\documentclass{JHEP}

\usepackage{epsfig}

\newcommand{\ra}{\rightarrow}
\def\lsim{\raise 0.4ex\hbox{$<$}\kern -0.8em\lower 0.62
ex\hbox{$\sim$}}
\def\gsim{\raise 0.4ex\hbox{$>$}\kern -0.7em\lower 0.62
ex\hbox{$\sim$}}
\newcommand\eq[1]{eq.~(\ref{#1})}
\newcommand\eqs[2]{eqs.~(\ref{#1}) and (\ref{#2})}
\newcommand\eqss[3]{eqs.~(\ref{#1}), (\ref{#2}) and (\ref{#3})}

\newcommand\eqst[2]{eqs.~(\ref{#1})--(\ref{#2})}
\newcommand\pa{\partial}
\newcommand\pdif[2]{\frac{\pa #1}{\pa #2}}
\newcommand\ee{\end{equation}}
\newcommand\be{\begin{equation}}
\newcommand\eea{\end{eqnarray}}
\newcommand\bea{\begin{eqnarray}}
\newcommand\sub[1]{_{\rm #1}}

\def\dslash{\not{\hbox{\kern-2pt $\partial$}}}
\def\Dslash{\not{\hbox{\kern-4pt $D$}}}
\def\pslash{\not{\hbox{\kern-2.3pt $p$}}}
\newcommand\gym{g_{\rm YM}}
\renewcommand\({\left(}
\renewcommand\){\right)}
\renewcommand\[{\left[}
\renewcommand\]{\right]}

\newcommand\fverb{\setbox\pippobox=\hbox\bgroup\verb}
\newcommand\fverbdo{\egroup\medskip\noindent%
                        \fbox{\unhbox\pippobox}\ }
\newcommand\fverbit{\egroup\item[\fbox{\unhbox\pippobox}]}
\newbox\pippobox

\title{Confinement, asymptotic freedom and renormalons
       in type 0 string duals }

\author{Roberto Grena, Simone Lelli, Michele Maggiore\thanks{presently
        at CERN, Theory Division, CH-1211, Geneva 23, Switzerland}
\hspace{1mm}  and Anna Rissone\\
        Dipartimento di Fisica, Universit\`a di Pisa and INFN,
        sezione di Pisa\\
        via Buonarroti 2, I-56127 Pisa, Italy\\}

\preprint{IFUP-TH 2000/15}        

\abstract{Type 0B string theory has been proposed as the dual
  description of non-supersymmetric  SU(N) Yang-Mills
  theory coupled to six scalars, in four dimensions. We study numerically and
  analytically the equations of motion of type 0B gravity
  and we find RG trajectories of the dual theory that flow from
  an asymptotically free UV regime to a confining
  IR regime. In the UV we find a one-parameter family of solutions
  that approach asymptotically $AdS_5\times S^5$ with a
  logarithmic flow of the coupling plus  non-perturbative
  terms that correctly reproduce all UV and IR renormalon singularities.
  The first UV renormalon gives a contribution $\sim F_1(E)/E^2$ and
  we are able to predict also the form of the function $F_1(E)$, which,
  from the YM side,  corresponds to summing all multiple-chain bubble
  graphs. The fact that the positions of the renormalon singularities
  in the Borel plane come out correctly is a non-trivial test of the
  conjectured duality.}

\begin{document} 


\section{Introduction and summary}

The attempts to generalize the Maldacena 
conjecture \cite{Mal1,GKP,Wit1,AGMOO} to four-dimensional
non-supersymmetric YM theories have followed different directions. 
The approach proposed by
Witten is to start from the duality between type II strings
and a five-dimensional super-YM; compactifying the fifth dimension on a circle
with different boundary conditions for bosons and fermions one breaks 
supersymmetry and is left with an effective four-dimensional 
non-supersymmetric theory~\cite{Wit2}. A conceptually similar 
possibility is to reduce the supersymmetry to $N=1$ or even 
$N=0$ with the addition of mass deformations~\cite{GPPZ1,GPPZ2,PS}. 

A different strategy is to start
directly from a string theory with worldsheet supersymmetry
but no spacetime supersymmetry. Polyakov has advocated this approach 
and has suggested the use of non-critical strings~\cite{Pol1,Pol2}, 
and Klebanov and Tseytlin \cite{KT1,KT2} have adapted this suggestion 
to critical type 0B string theory. On the gravity side one considers a
configuration of $N$ electric D3-brane in the 
type 0B theory, and the dual field theory is
conjectured to be $SU(N)$ four-dimensional YM plus six adjoint
scalars. The six scalars are the price to pay for working in the 
critical dimension.

In this paper we follow the latter approach and we study the
renormalization group (RG) flow derived from the equations of motion of
type 0B gravity.
Klebanov and Tseytlin~\cite{KT2} 
have found that in the UV one recovers asymptotic
freedom,
\be\label{log}
\frac{1}{\gym^2(E)}\sim \ln \frac{E}{\Lambda}-
{\rm const.}\, \ln\ln \frac{E}{\Lambda}\, .
\ee
The numerical value of the beta function coefficients are not under
control because of various corrections discussed below, but
it is very encouraging to see the correct logarithmic behaviour.
Minahan~\cite{Min1,Min2} has instead presented confining IR
solutions. Further related work appeared in
refs.~\cite{FM}-\cite{AFS}. It
is therefore natural to ask whether asymptotic freedom and confinement
are connected by a single RG flow.

In this paper we perform a detailed analytic and numerical
investigation of the RG flow, and we find the following
results.

\begin{itemize}

\item The UV solution found in refs.~\cite{KT2,Min2} is just a member of a
one-parameter family of solutions. The free parameter $\kappa$, which is
related to subleading terms in the UV,  determines the IR
behaviour of the solution. For $\kappa$ smaller than a critical value
$\kappa_c$ (that we can determine analytically) the solution flows in the IR
toward a confining solution. The case considered in \cite{KT2}
corresponds instead to $\kappa >\kappa_c$, and 
it was found in \cite{KT2} that
in this case the solution flows in the IR toward a conformal fixed
point at infinite coupling.

\item We will then show, both numerically and analytically, the existence
of RG trajectories that connect asymptotic freedom in the UV with
confinement in the IR.

\item We will find that in the UV, beside the logarithmic terms shown
in \eq{log}, type 0B gravity  predicts  also power corrections,
\be
\frac{1}{\gym^2(E)}\sim \ln \frac{E}{\Lambda}-
c_1\ln\ln \frac{E}{\Lambda}
+c_2F_1(E) \frac{\Lambda^2}{E^2} +c_3F_2(E) \frac{\Lambda^4}{E^4} +
c_4 \frac{\Lambda^4}{E^4}+
O(\frac{1}{E^6})
\ee
where the $c_i$ are constants to be defined below.
These power corrections match exactly the non-perturbative
contributions coming from renormalons. The term
$\sim F_1(E)/E^2$ on the field theory side comes from the first UV
renormalon, the term $\sim F_2(E)/E^4$ from the second UV renormalon,
and similarly we get  the contributions of all UV
renormalons. The term $c_4/E^4$ is the leading IR renormalon, and we
correctly find no IR renormalon contribution $1/E^2$. Thus, we
get the right positions of all renormalon singularities in the
Borel plane.

\item Even the energy dependence of the prefactors, that is
$F_1(E)$ for the first
UV renormalon and a constant for the leading IR renormalon, matches
what is known from the field theory side. In particular, we will
predict the energy dependence
\be
F_1(E)\sim  \sin\( \beta_T\ln \frac{E}{\Lambda}+\alpha_T \) 
\ee
with $\alpha_T,\beta_T$ constants. Expanding at first order in
$\beta_T$ we get a result consistent with what is obtained
in QCD summing diagrams
with a single bubble chain. The above closed form is a prediction of
the dual string theory for the resummation of multiple chain bubble
graphs, which would be very interesting to check from the field theory
side. 

\end{itemize}

The above results hold at lowest order in $\alpha '$ and string
loop corrections, and are presented in sects.~3-6. The fact that
some features of the UV limit of the YM theory can be obtained 
from the gravity side is surprising, since the 
$SU(N)$ YM and the gravity descriptions are duals to each other 
and are based  on opposite expansions, in $g^2N$ and $1/\sqrt{g^2N}$,
respectively. It is therefore important to study the corrections to
the gravity predictions in the UV limit. We will discuss them in
sect.~7. It has been found in ref.~\cite{KT2} that, if the $\alpha'$
corrections can be expressed in terms of the Weyl tensor, then the
only price to pay in the UV is that the numerical values of the beta
function coefficients cannot be predicted, but the logarithmic running
of the coupling is unaltered. Furthermore we will find that, under
the same conditions, the
position of the renormalon singularities in the Borel plane is
independent of the $\alpha'$ corrections.

\section{General strategy for computing the RG flow}
In this section we discuss the general strategy and the assumptions
used to extract the RG
flow in the dual theory from the gravity equations of motion of type 0
strings. The low energy effective action of type 0B strings~\cite{DH,SW,Pol} 
is written in terms of a (closed string) 
tachyon field $T$, the dilaton $\phi$, the metric, and a
doubled set of RR fields.
The terms involving the tachyon in the
effective action have been discussed by Klebanov and
Tseytlin~\cite{KT1}, and are
\be\label{act}
\int d^{10}x\sqrt{G}e^{-2\phi}\( 
\frac{1}{2}G^{mn}\pa_mT\pa_nT +V(T) \)+
\int d^{10}x\sqrt{G} \, \frac{1}{2}f(T)|F_5|^2\, .
\ee
Because of world-sheet supersymmetry, the tachyon potential $V(T)$
has only even powers of $T$. We write (to have the same notations
of~\cite{Min2}) $V(T)=-2g(T)$, and
\be
g(T)=-\frac{1}{4}m^2T^2-\lambda T^4+O(T^6)
=\frac{1}{2}T^2 +O(T^4)\, .
\ee
The tachyon mass is $m^2=-2/\alpha'$ and we have set $\alpha'=1$.
The function $f(T)$ describes the coupling of the tachyon to the 
5-form field strength $F_5$, and is given by
\be\label{f(T)}
f(T)=1+T+\frac{1}{2}T^2+O(T^3)
\ee
and  the coupling to the other RR fields can be found in ref.~\cite{KT1}. Note
that, due to the doubling of RR fields, the 5-form field strength is
not constrained to be self-dual, so that $|F_5|^2$ is non
vanishing. D-branes in this theory have been studied
in~\cite{BG,KT1}. Another consequence of the doubling of the RR fields
is that there are
two kinds of D-branes, electrically and magnetically charged.
It is then interesting to consider the solution of the gravity 
equations of motion corresponding to a large number $N$ of coinciding 
electrically charged 3-branes, since from \eq{act} we see that
the coupling of the 3-brane to
the 5-form RR field strength can stabilize the closed string
tachyon. The tachyon from open strings which ends on
parallel D-branes is instead eliminated by the GSO 
projection~\cite{KT1,Pol1}. 
In the Einstein frame, the ansatz for the 3-brane metric can be conveniently
written in terms of two functions $\xi$ and $\eta$ 
of a transverse variable $\rho$,
\be\label{metrica}
ds_E^2=e^{\frac{\xi}{2}-5\eta}d\rho^2+e^{-\frac{\xi}{2}} dx_{(4)}^2+
e^{\frac{\xi}{2}-\eta} d\Omega_{(5)}^2\, .
\ee                
For the 4-form RR field $C_{(4)}$ one takes  $C_{0123}(\rho )$ as the only
non-vanishing component, so that $F_5=dC_{(4)}$ has
only the component $(F_{5})_{0123\rho}$;
similarly one takes $T=T(\rho ),\phi =\phi (\rho )$. The
equation of motion for the RR field can be explicitly integrated and
gives an integration constant $Q$ which is just the RR charge, and is
proportional to $N$. The
remaining equations of motion then separate into four dynamical
equations~\cite{KT1,Min2} 
\bea
\label{a}\ddot\phi+\frac{1}{2} g(T)
e^{\frac{\phi+\xi}{2}-5\eta}=0\, ,  \\
\label{b}\ddot\xi+\frac{1}{2}g(T) e^{\frac{\phi+\xi}{2}-5\eta}+
\frac{2Q^2}{f(T)} e^{-2\xi}=0\, , \\
\label{c}\ddot\eta+\frac{1}{2}g(T) e^{\frac{\phi+\xi}{2}-5\eta}+
8 e^{-4\eta}=0 \, , \\
\label{d}\ddot T+\frac{2Q^2 f'(T)}{f^2(T)} e^{-2\xi}+2 g'(T)
e^{\frac{\phi+\xi}{2}-5\eta}=0\, ,
\eea 
and (due to the invariance under reparametrizations of $\rho$)
a constraint on the initial values, conserved by the dynamical
equations, 
\be\label{vincolo}
\frac{{\dot\phi}^2}{2}+\frac{{\dot\xi}^2}{2}-5{\dot\eta}^2+
 \frac{{\dot T}^2}{4}+g(T) e^{\frac{\phi+\xi}{2}-5\eta}+20
e^{-4\eta}-
Q^2\frac{e^{-2\xi}}{f(T)}=0\, .
\ee
The dot is the derivative with respect to $\rho$.
The explicit form of $f(T)$ plays an
important role  in the UV regime, where the tachyon is stabilized at its
minimum, $f'(T_0)=0$~\cite{KT1}.

There are uncertainties due  to the fact that neither $f(T)$ nor
$g(T)$ are known in closed form; actually the latter is not
even unambiguously defined, since, using the equations of motions, 
a term $\sim\nabla^2T$ in the effective action can be traded for a
term $\sim m^2T$ so that, for
instance, a higher derivative 
term $\sim (\nabla^2 T )^4$ in the effective action
can be reabsorbed into a $\sim T^4$ term.
In the following we will at first  set $g(T)= (1/2)T^2$
and $f(T)=1+T+T^2/2$ (the minimum of $f(T)$ is in this case at $T_0=-1$).
We will then discuss how our results depend on these choices.

Our aim is to solve \eqst{a}{vincolo}, and to translate the solutions 
into a renormalization group
flow of the coupling $\gym^2$ of the dual theory. The first issue is
how to read $\gym^2$ from the gravity side. We follow the approach of
\cite{Min1,Min2,KT2} and we read it from the quark-antiquark potential,
computed from the Wilson loops as in refs.~\cite{Mal2,RY}. 
This means that we take
\be\label{gym}
\gym^2\sim \exp\{\phi (\rho )/2\}
\ee
as our definition of the coupling. The exact proportionality factor
will not concern us here since, as found in refs.~\cite{Min2,KT2} and as we
will discuss below, the gravity solution is in any case
subject to a number of uncertainties, due to $\alpha'$ corrections and
to the exact form of  $f(T),g(T)$, that affect the numerical
value of the proportionality coefficient in \eq{gym}.

The definition (\ref{gym}) is apparently in contradiction with what
one would obtain from the D-brane effective action, which seems to
give $\gym^2\sim \exp\{\phi\}$ rather than $\sim\exp\{\phi
/2\}$. However, as discussed in \cite{KT2}, in the type 0 theory there
exists also a tachyon tadpole on the D-brane, so that the effective
action of the D-brane is proportional not simply to $e^{-\phi}$ but
rather to $k(T)e^{-\phi}$, with $k(T)=1+T/4+O(T^2)$. Since
in the UV the tachyon runs toward its minimum at $T_0=-1$, higher
powers in $k(T)$ cannot be neglected and the closed form of $k(T)$ is
needed to draw conclusions. It is clear, however, that this is a point
which  deserves further investigations.

Eq.~(\ref{gym}) provides us with the dependence of the coupling on the
transverse coordinate $\rho$. To obtain an RG flow, the second step is
to connect $\rho$ with a physical energy scale. This can be done as
follows. From \eq{metrica} we see that a dilatation of the
4-dimensional coordinates, $x\ra\lambda x$, can be reabsorbed into a
rescaling of $\xi$ such that $\exp\{-\xi /2\}$ scales like an energy
squared (while rescaling $\eta$ and $\rho$ so to keep the other terms
invariant). We will then define the energy scale from
$\exp\{-\xi (\rho )/2\}\sim E^2$, or
\be\label{E}
\ln \frac{E}{\Lambda}=-\frac{1}{4}\xi (\rho )\, .
\ee
$\Lambda$ is a scale that we will determine later.
In the UV region the solutions will approach $AdS_5\times S^5$
and, writing the  metric of $AdS_5$ as $\sim du^2/u^2 +u^2dx_{(4)}^2$, we 
see that $\exp (-\xi /2)\sim u^2$ and then
the prescription (\ref{E}) reduces to the by now standard identification
$u\sim E$. Far from the UV there is a certain arbitrariness in this
definition, that we regard as corresponding to the arbitrariness in
the choice of the renormalization scheme on the gauge theory side. 
Of course the details of the
beta function are not independent of these choices, and only
universal properties, like the existence of zeros and how the zeros
are approached are what really matters.

Solving the equations of motion we  get $\phi (\rho)$ and $\xi
(\rho)$. Combining \eqs{gym}{E} then provides the dependence of 
$\gym^2$ from the energy scale and therefore the RG flow.

\section{The UV and IR asymptotics}\label{asy}
An asymptotic solution valid in the UV limit has been presented 
in ref.~\cite{Min2} and, as a more systematic expansion, in
ref.~\cite{KT2}. Direct inspection of \eqst{a}{vincolo}
reveals however that there is actually a one-parameter family of solutions. 
It is convenient to introduce a new variable $y$ from
\be\label{rho}
\rho \equiv e^{-y}\, .
\ee
In the limit $y\ra\infty$ the solution  is
\bea
\label{phi}
\phi &=&-2\ln y+15\ln 2+\frac{1}{y}(39 \ln y +\kappa )+
O\(\frac{\ln^2 y}{y^2}\)\, ,\\
\label{xi}
\xi &=&-y+\ln 2 +\frac{1}{y}+
\frac{1}{2y^2}(39\ln y+\kappa -104)+O\(\frac{\ln^2y}{y^3}\)\, ,\\
\label{eta}
\eta &=&-\frac{y}{2} +\ln 2 +\frac{1}{y}+
\frac{1}{2y^2}(39\ln y +\kappa -38)+O\(\frac{\ln^2y}{y^3}\)\, ,\\
\label{T}
T&=&-1+\frac{8}{y}+\frac{4}{y^2}(39\ln y +\kappa -20)+
O\(\frac{\ln^2y}{y^3}\)\, .
\eea
Here and in the following
we  set $Q=1$; we see from the equations that
the solution for generic $Q$ can be recovered 
from 
\bea\label{Q}
\xi (y;Q) &=& \xi (y;Q=1)+\ln Q\, ,\nonumber\\ 
\phi (y;Q)&=& \phi (y;Q=1) -\ln Q\, ,
\eea
while $\eta ,T$ are independent of $Q$; $\kappa$ is a free parameter, and
the solution found in ref.~\cite{KT2} corresponds to $\kappa =0$. In spite
of the fact that in the UV $\kappa$ only appears in terms which look
quite subleading, we will find that its value is very
important for determining the IR behaviour of the 
solution.\footnote{This free parameter comes out because, inserting 
  into the equations of motion an
  ansatz of the form of \eqst{phi}{T} with generic coefficients, 
  one finds that in \eq{a}
  the terms of order $(\ln y)/y^2$ cancel automatically, and
  therefore impose no constraint on the coefficients of the solution.
  Looking only to the terms up to $O(1/y)$ one might think that
  this parameter can be removed by a conformal rescaling, but this is
  not true anymore when one considers 
  also  the terms $O(1/y^2)$ in $\xi ,\eta$.}
In the large $y$ limit $\xi\sim -y\ra -\infty$. Eq.~(\ref{E}) then 
shows that $y\ra\infty$ (or $\rho\ra 0$)
is the UV region. At $y=\infty$ the metric reduces
to $AdS_5\times S^5$.

Using the prescription discussed in sect.~2, one can now extract the RG
flow in the UV, with the result~\cite{KT2}
\be\label{uv}
\gym^2(E)\sim\frac{1}{\ln (E/\Lambda )-\frac{39}{8}\ln\ln (E/\Lambda )}\, .
\ee
Note that the logarithmic terms are independent of $\kappa$.
Eq.~(\ref{uv}) should be compared to the running of the coupling in the proposed
dual theory, that is four-dimensional $SU(N)$ YM theory with 6 scalars in the adjoint
representation, which at the two-loop level is
\be\label{2loop}
\gym^2(E)=\frac{8\pi^2}{b_1\(\ln (E/\Lambda )
+\frac{b_2}{2b_1^2}\ln\ln (E/\Lambda )\)}
\ee
with
\be\label{beta0}
b_1=\frac{8}{3}N\, , \hspace{15mm} 
\frac{b_2}{2b_1^2}=-\frac{3}{16}\, .
\ee
We see that \eq{uv} qualitatively reproduces the logarithmic running of the coupling in
the UV.  The precise value of the beta function coefficients cannot be
reliably estimated, since the gravity prediction is affected  by $\alpha'$
corrections (see  refs.~\cite{KT2,Min2} and sect.~\ref{alpha'}). 
Furthermore, with a generic tachyon potential $g(T)$, and
a generic function $f(T)$ with a minimum at $T=T_0$, the factor 
$39/8$ in \eq{uv} is modified as~\cite{Min2}
\be
\frac{39}{8}\ra \frac{7}{8}+\( \frac{g'(T_0)}{g(T_0)} \)^2\, ,
\ee
so it is clear that the values of the numerical coefficients
are not under control.

In the opposite limit, $\rho\ra \infty$, 
it is again possible to find
asymptotic  solutions. In this case however
there is a much larger variety of solutions. In particular
Minahan~\cite{Min2} has considered the generic behaviour
\bea
\phi & \simeq &\phi_1\rho+\phi_0\,\nonumber \\
\xi  & \simeq &\xi_1\rho+\xi_0\, \label{conf}   \\
\eta & \simeq &\eta_1\rho+\eta_0\,\nonumber \\
T    & \simeq &t_1\rho+t_0\, ,\nonumber 
\eea
At large $\rho$ this is a solution of the equations of motion if
\be\label{disu}
\xi_1,\eta_1 > 0\, ,\hspace{5mm}
5\eta_1-\frac{1}{2}\phi_1-\frac{1}{2}\xi_1>0
\ee
and
\be\label{isu}
\frac{1}{2}\phi_1^2+\frac{1}{2}\xi_1^2-5\eta_1^2+
\frac{1}{4}t_1^2=0\, .
\ee
The latter equation follows from the constraint equation
(\ref{vincolo}), while the inequalities (\ref{disu}) ensure that all exponentials
in \eqst{a}{vincolo} are suppressed and therefore that the equations of
motions are satisfied.

In the limit $\rho\ra\infty$ we now 
have $\xi\ra+\infty$ and \eq{E} tells that
this is the IR limit. To investigate confinement 
one can  compute the quark-antiquark potential as
in ref.~\cite{Mal2,RY,RTY,BISY}. One considers a Wilson loop on the boundary
$\rho =0$, with edges along the directions $x,t$,
and looks for 
the classical string world-sheet $\rho (x,t)$ which has
the Wilson loop as its boundary. The Nambu-Goto action is
\be\label{NG}
S_{NG}=\frac{1}{2\pi\alpha '}\int d\sigma d\tau 
\sqrt{
\det (G_{MN}\pa_{\alpha}X^M\pa_{\beta}X^N)}\, ,
\ee
where $G_{MN}$ is the string frame metric, which in ten dimensions is
related to the Einstein frame metric $G_{MN}^{(E)}$ by
$G_{MN}=e^{\phi /2}G_{MN}^{(E)}$, and $G_{MN}^{(E)}$ for the 3-brane
solution is given in~\eq{metrica}. Inserting $G_{MN}$ into \eq{NG} 
one finds that  the static potential $V(L)$ of a $q\bar{q}$ pair
separated by a distance $L$ along the direction $x$ is  obtained 
from~\cite{Min2}
\be
V(L)=\frac{1}{2\pi\alpha'}\int_0^L dx\, 
\[ e^{\phi-5\eta} \( \pdif{\rho}{x} \)^2 + e^{\phi-\xi} \]^{1/2}\, ,
\ee
after subtracting the divergent part corresponding to the energy of two
separated massive quarks~\cite{Mal2,AGMOO}. 
The above result holds for
generic $Q$. The $Q$ dependence can be extracted using \eq{Q} and
gives
\be
V(L)=\frac{1}{2\pi\alpha'\sqrt{Q}}\int_0^L dx\, 
\[ [e^{\phi-5\eta}]_{Q=1}
 \( \pdif{\rho}{x} \)^2 + 
\frac{1}{Q} [e^{\phi-\xi}]_{Q=1} \]^{1/2}\, .
\ee
The result  depends crucially on  the function
\be
\tilde{f}^2(\rho )\equiv [e^{\phi-\xi }]_{Q=1}\,
\stackrel{\rho\ra \infty}{\longrightarrow}
e^{ (\phi_1-\xi_1)\rho  +(\phi_0 -\xi_0)}\, .
\ee
(We define $\phi_0,\xi_0$ as the coefficient of the solution with $Q=1$).
If $\phi_1<\xi_1$, $\tilde{f}$ vanishes at $\rho =\infty$, and
the classical string configuration will go all the
way to the region where $\rho\ra \infty$, 
where there is no cost in energy in separating the
quarks further; for $\phi_1=\xi_1$, the classical string
configuration  again goes to the minimum of $\tilde{f}(\rho )$ at
$\rho\ra\infty$, but now $\tilde{f}(\infty )=
\exp\{(\phi_0 -\xi_0) \}$ is non-vanishing; the 
$q\bar{q}$ potential at large $L$ then becomes
\be
V(L)\simeq\sigma L\, ,
\ee
with 
\be\label{tension}
\sigma =\frac{1}{2\pi\alpha'Q}\, e^{ (\phi_0 -\xi_0)/2 }
\ee
If instead $\phi_1 > \xi_1$, then the function $\tilde{f}(\rho )$ diverges at
$\rho\ra\infty$. On the other hand, it also diverges in the UV, where
$(\phi -\xi )\ra y\ra +\infty$, see \eqst{phi}{xi}.
This means that it has a minimum at some value $\rho\sub{min}$, where
the classical string configuration goes. Then we have again a linear
$q\bar{q}$ potential with 
\be\label{tension2}
\sigma =\frac{1}{2\pi\alpha'Q}\,\tilde{f}(\rho\sub{min};Q=1 )\, .
\ee
This gives the relation between the `QCD' string tension 
$\sigma$,
the type 0 string tension $1/(2\pi\alpha ')$ and the RR charge $Q$.
In the case of $SU(3)$ YM, the experiment gives $\sigma\sim
(1\, {\rm fm})^{-2}$ and eqs.~(\ref{tension}) or (\ref{tension2}) fix the value of 
$\alpha '$ for the dual string theory.

Summarizing, the condition for confinement is 
\be\label{cond}
\phi_1 \geq \xi_1\, .
\ee

A very different IR solution has been found by Klebanov and
Tseytlin~\cite{KT2}. In this case, approaching the IR limit, the
Einstein frame metric becomes again asymptotic to $AdS_5\times S^5$,
while the dilaton diverges and the tachyon goes to zero. 
Therefore the dual theory flows toward
a conformally invariant point with infinite coupling.

It is at this point natural to ask whether the one-parameter family of
UV solutions discussed before is smoothly connected to any of these
IR solutions. This is the issue that we will address in the next
section numerically and in sect.~\ref{sec4} analytically.

\section{Numerical integration}\label{nume}
The numerical study of \eqst{a}{d} is not as straightforward as one
might hope; actually we found that, if we start from the UV with
initial conditions that reproduce the asymptotic behaviour given by
\eqst{phi}{T}, the numerical integration runs almost
immediately into divergencies. The reason for this numerical problem will
become clear at the end of this section, but
we have found that starting instead from the IR with a solution of
the type (\ref{conf}) and integrating toward the UV poses no numerical
problem.

However, this
class of IR solutions from which we are starting the integration
is characterized by more free parameters than
the  solution (\ref{phi})-(\ref{T}) to which  we would like to connect
in the UV. This means that we have to scan the parameter space of the
IR solutions in order to find those very particular solutions (if any)
that match to the desired UV behaviour. This can be done as
follows. We start by studying an IR ($\rho\gg 1$) solution of the
asymptotic form
\be
\phi (\rho )=\rho\, ,\hspace{5mm}\xi (\rho)=\rho \, ,\hspace{5mm}
\eta (\rho )=\frac{1}{\sqrt{5}}\rho +\eta_0\, ,\hspace{5mm}T(\rho )=0\, ,
\ee
keeping $\eta_0$ as the only free parameter. This is a solution of the
type (\ref{conf}), where we have chosen
$\phi_1=\xi_1$ in order to have a confining solution, see
\eq{cond}, and
we have arbitrarily set $\xi_1=1$. We also specialize 
to solutions
such that in the IR the tachyon goes to zero.\footnote{We 
have also studied  the case of a tachyon linearly growing 
in the IR, i.e. $t_1\neq 0$ in \eq{conf}, 
and we have found that in the UV the behaviour of the
solution is essentially the same as in the case $t_1=0$. However, the case
$t_1=0$ is in a sense more solid because  our
ignorance on higher orders in $T$ of the functions $f(T), g(T)$ becomes 
irrelevant in the IR.} Eq.~(\ref{isu}) then
fixes $\eta_1=1/\sqrt{5}$. The inequalities (\ref{disu}) are satisfied.
The invariance under constant shifts in $\rho$ allows to fix, e.g.,
$\xi_0=0$. We also  set 
$\phi_0=0$; from \eq{tension} we see that a different value of $\phi_0$
reflects itself on a quantitatively different value of 
the string tension $\sigma$, but
we expect that this does not give rise to qualitative differences in
the solution (as we have indeed checked numerically, setting
$\eta_0=0$ and using $\xi_0$ as a free parameter).

The results of the numerical integration are as follows. For $\eta_0$
smaller than a critical value $\eta_c$, we find that
$\eta (\rho )\ra -\infty$ at a finite value of 
$\rho =\rho_0$, which
therefore sets a limit to the range  of $\rho$. 
The fact that $\rho$ cannot range from $-\infty$ to $+\infty$ 
is to be expected. In the usual Dp-brane solutions $\rho$
is indeed defined only in a semi-infinite range
which, after a constant shift of $\rho$, can always 
be taken to be $0\leq \rho <\infty$, and $\rho =0$ is the D-brane horizon.
However, in our case we find that  $\xi (\rho_0)$ 
is still finite. Since $\xi$ is
related to the energy scale, $\xi =-4\ln (E/\Lambda )$, this means that in these
solutions 
there is no region that we can interpret as an UV limit of the dual theory
(or, in the gravity language,  these solutions have no horizon). Therefore the
solutions that start in the IR with $\eta_0 < \eta_c$ are not connected
to any of the UV solutions given by
\eqst{phi}{T}.

If instead  $\eta_0 > \eta_c$, the situation is reversed and 
$\xi\ra -\infty$ 
at a finite value  $\rho_0$ while $\eta (\rho_0)$ and $\phi (\rho_0)$ are still
finite. This means that at $\rho_0$ the energy scale
$E\ra\infty$, so now the solution reaches the UV regime. However,
$\phi(\rho_0)$ is finite and therefore $\gym^2$ does not go to
zero in the UV. Then, again, this IR solution does not match to the
desired UV solution.

The situation is however different for $\eta_0=\eta_c$. In this case
$\phi,\xi$ and $\eta$ all go toward $-\infty$ at the same value 
$\rho =\rho_0$, 
while $T\ra  -1$; therefore, there are chances of
matching this solution with the UV solution of \eqst{phi}{T}. 

It is convenient to introduce the variable $y$ in analogy to
\eq{rho}, so that the UV corresponds to $y\ra\infty$. In the analytic
computation of sect.~(\ref{asy}) the UV limit was at $\rho =0$ while
now it is at $\rho =\rho_0$, and to have the same convention we must
make a constant shift in $\rho$ by $- \rho_0$, so we
define $y$ from
\be
\rho -\rho_0 = e^{-y}\, .
\ee
With repeated runs we
have located accurately the value of $\eta_c$ and the corresponding
$\rho_0$, obtaining $\eta_c=0.4962700(1),
\rho_0=-0.2076889(1)$. Locating $\eta_c$ with great precision is
of course necessary if we want to follow the solution deep into the UV
regime. With a precision $O(10^{-7})$ on $\eta_c$ we  estimate
that we can reliably follow the solution in the UV region up to 
$y\sim 7\ln 10=O(10)$. 

\begin{figure}
\centering
\includegraphics[width=0.7\linewidth,angle=0]{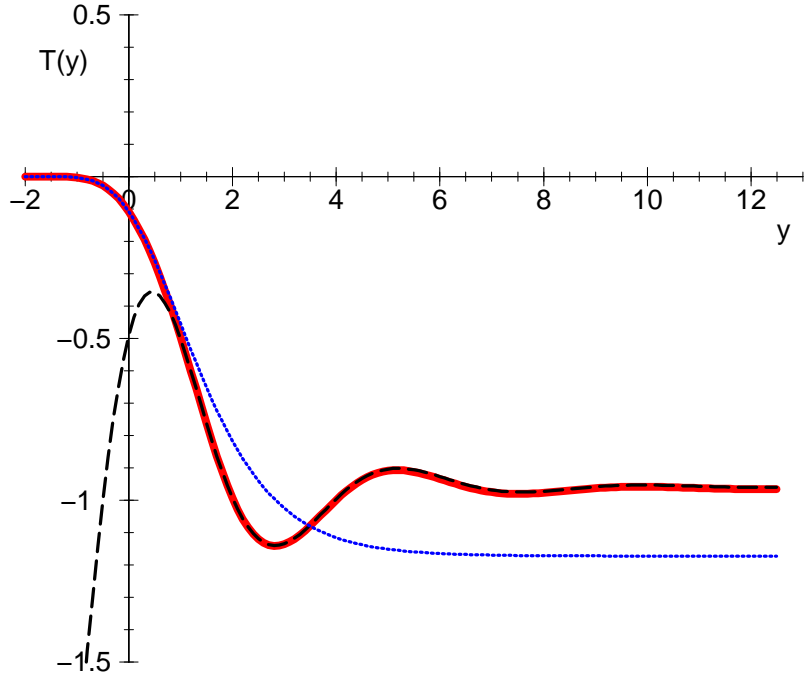}
\caption{The result of the numerical integration for
  $T (y)$ (red solid line) discussed in sect.~(\ref{nume}),
together with the analytic results in
  the UV (black, dashed) and
  in the IR (blue, dotted) discussed in sect.~\ref{sec4}. }
\label{figT}
\end{figure}

\begin{figure}
\centering
\includegraphics[width=0.7\linewidth,angle=0]{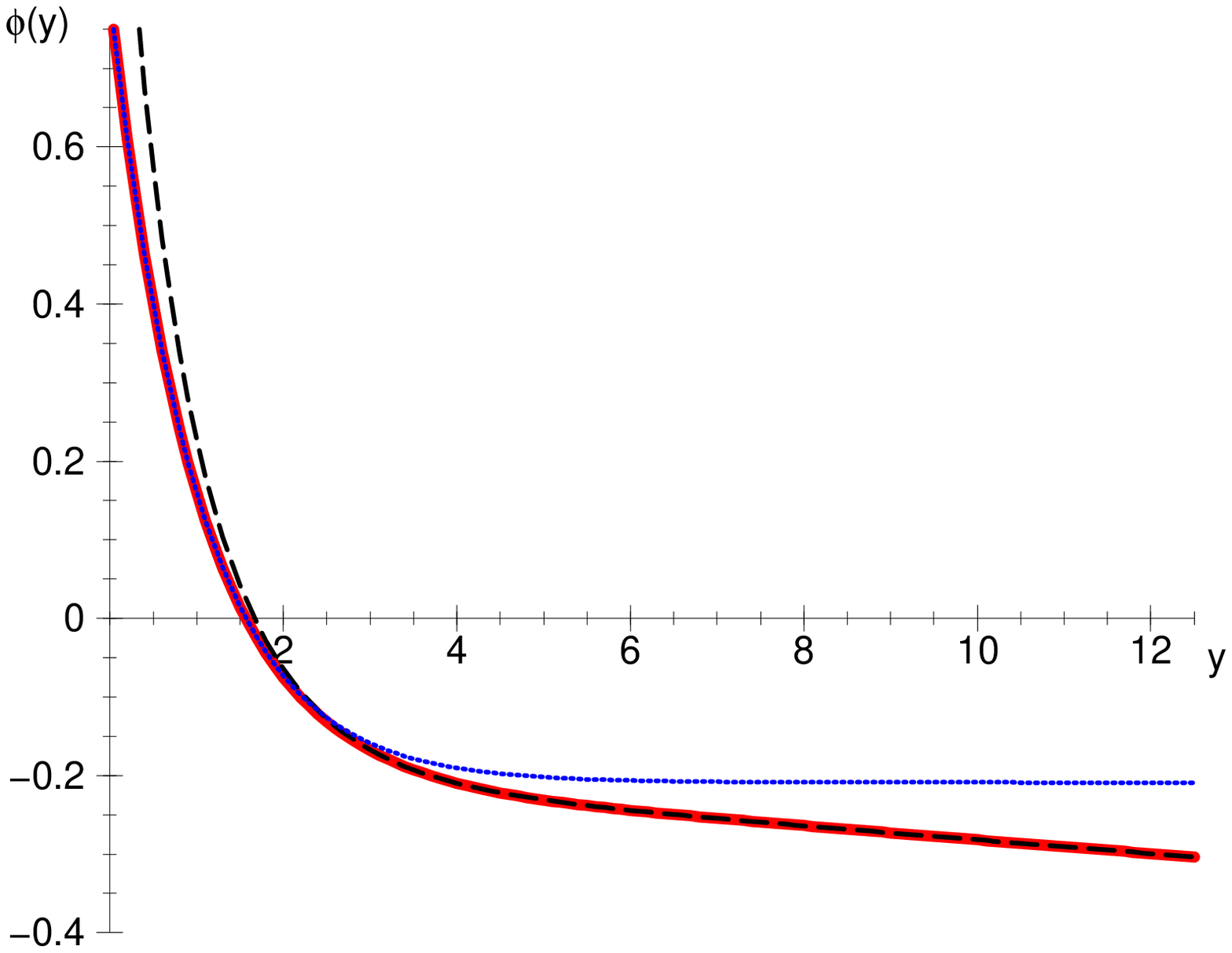}
\caption{The result of the numerical integration for
  $\phi (y)$ (red solid line)  discussed in sect.~(\ref{nume}),
 together with the analytic results in
the UV (black, dashed) and
  in the IR (blue, dotted) discussed in sect.~\ref{sec4}. }
\label{figphi}
\end{figure}

\begin{figure}
\centering
\includegraphics[width=0.7\linewidth,angle=0]{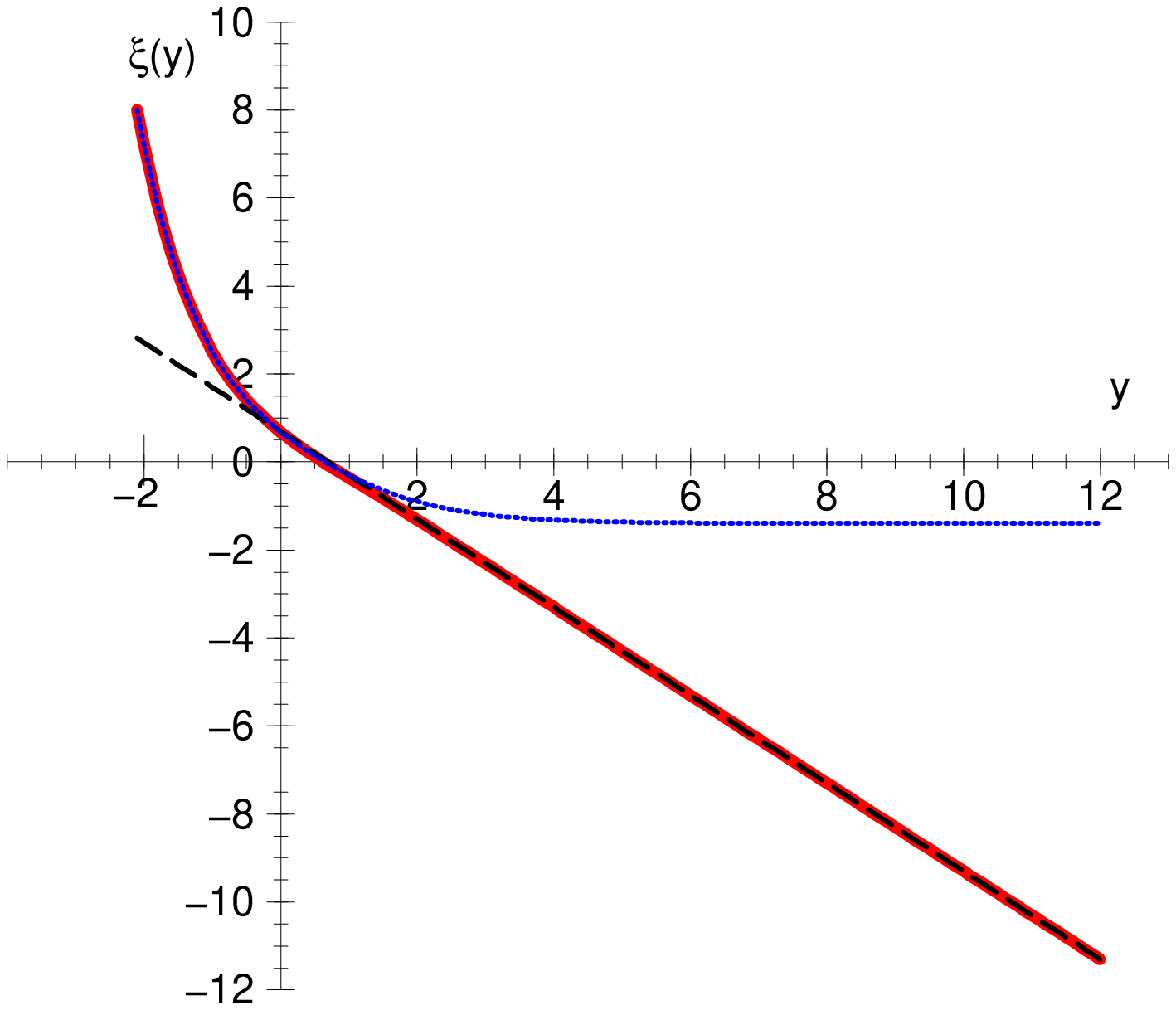}
\caption{The result of the numerical integration for
  $\xi (y)$ (red solid line)  discussed in sect.~(\ref{nume}),
 together with the analytic results in
the UV (black, dashed) and
  in the IR (blue, dotted) discussed in sect.~\ref{sec4}. }
\label{figxi}
\end{figure}

\begin{figure}
\centering
\includegraphics[width=0.7\linewidth,angle=0]{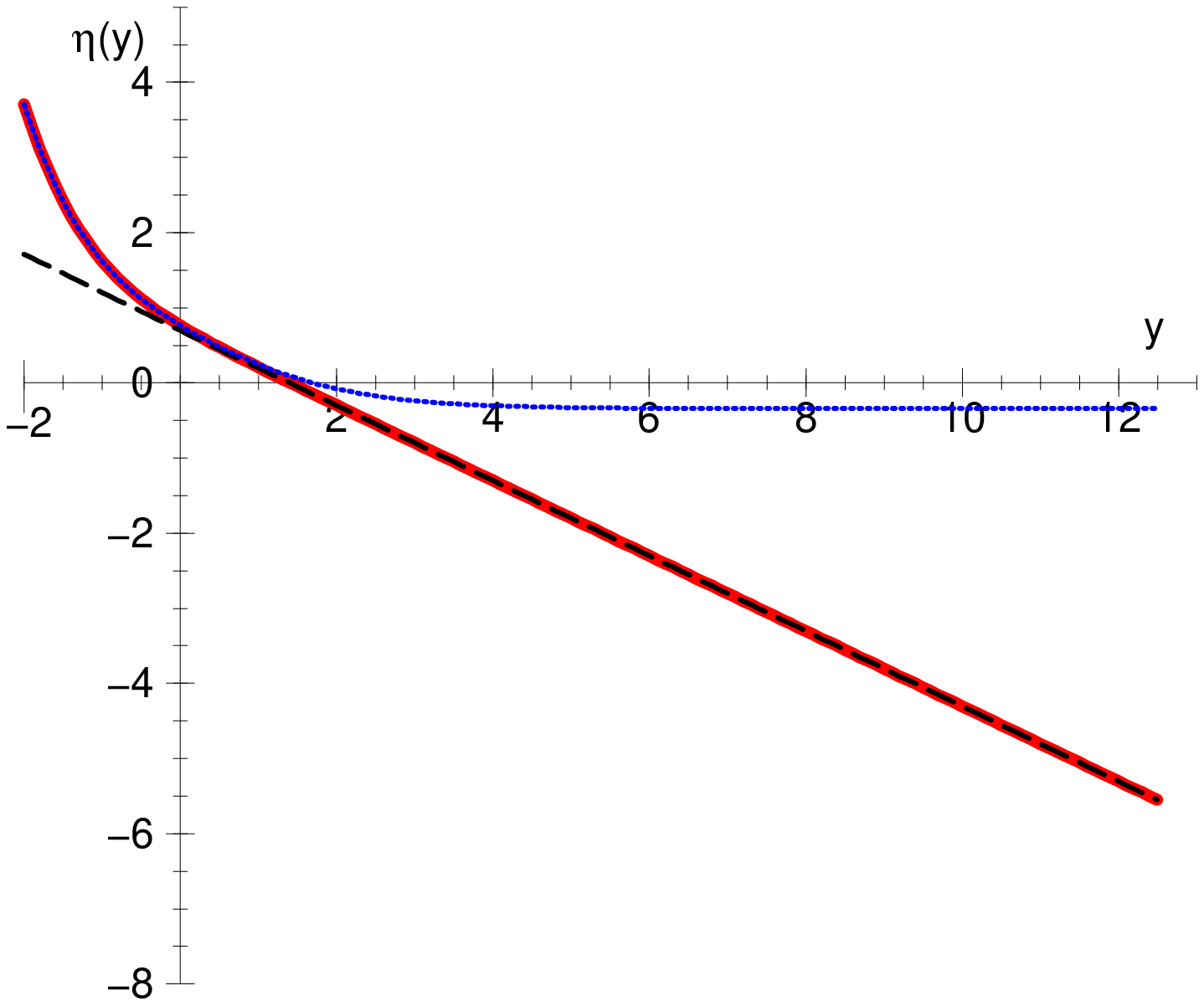}
\caption{The result of the numerical integration for
  $\eta (y)$ (red solid line)  discussed in sect.~(\ref{nume}),
 together with the analytic results in
the UV (black, dashed) and
  in the IR (blue, dotted) discussed in sect.~\ref{sec4}. }
\label{figeta}
\end{figure}

\begin{figure}
\centering
\includegraphics[width=0.7\linewidth,angle=0]{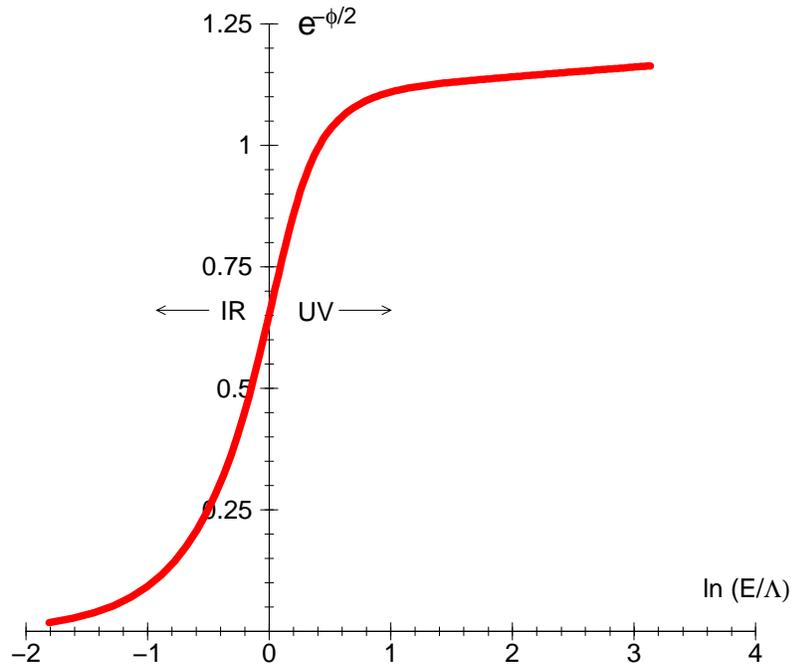}
\caption{$e^{-\phi /2}$ (which is proportional to
$1/\gym^2$) against $-\xi /4 =\ln (E/\Lambda )$. }
\label{figgym}
\end{figure}

The solution is shown in figs.~(\ref{figT})-(\ref{figeta}) (solid lines), plotted
against $y$.
The functions $\xi ,\eta$ and $T$ in the UV are very well fitted by the
leading terms of \eqss{xi}{eta}{T}, suggesting that we have succeeded
in matching the confining IR solution to this solution.
For the dilaton, however, the situation is more subtle.
Fig.~(\ref{figgym}) shows $e^{-\phi /2}\sim 1/\gym^2$ plotted against
$-\xi /4$, i.e. against $\ln (E/\Lambda )$. 
In the IR the coupling is strong and 
$\gym^2\sim 1/E^2$, while in the UV regime $1/\gym^2$ scales 
linearly with $\ln E$.

Thus, first of all we see that we indeed succeeded in matching a 
confining IR solution with an asymptotically free UV regime, and now
we want to understand whether this UV solution is related to \eq{phi}.
If one  compares with the data shown in
fig.~(\ref{figphi}) one finds that, numerically, 
the expansion given by \eq{phi} misses
badly. It is instructive to understand the reason.
In the UV region the numerical result is very well fitted by
\be
e^{-\phi/2} \simeq A+by +O(\ln y)\, ,
\ee
and the fit gives  $b\simeq 0.005$. 
Thus, the solution in the UV is well reproduced by
\be\label{rep}
\phi \simeq -2\ln\( A+b y +O(\ln y)\)\, .
\ee
At asymptotically large values of $y$, this is the same as
\be\label{expa}
\phi \simeq -2\ln y -2\ln b +O(\ln y/y) +O(1/y) \, .
\ee
This is just the leading behaviour predicted by \eq{phi}, which also
predicts $b=2^{-15/2}\simeq 0.0055$, in excellent agreement with the 
value from the fit. 

However, the expansion  of $\ln (A+b y)$ is  valid only
if $by\gg A$, and here  $A$
is of order one while $b\simeq 0.0055$ is quite small. So, while \eq{rep}
reproduces the data very well, its expansion, and therefore \eq{phi}, is 
only valid  for $y\gg 1/b\simeq 181$, which is well beyond the point
where we can push the numerical integration\footnote{This also 
  explains why we failed to integrate the solution
  starting from the UV asymptotics: we gave initial conditions that
  would have reproduced \eqst{phi}{T} in a region $y\sim 10$ where
  this is not a good approximation to the solution. On the other hand,
  in the numerical integration it is very difficult to start from
  much higher values of $y$, because
  one is confronted with exponentially small terms in the equations of
  motion.}.
 Then, it should be clear that
our numerical solution in the UV is indeed nothing but a member of the 
family of solutions given in \eqst{phi}{T}, and that the analytic
expansion for $\phi$ given in \eq{phi} 
is only valid for $y\gg 181$, rather than for $y\gg 1$. However, to clear up
any doubt, in the next section we will present an improved
analytic solution of the equations
that is valid for $y\gg 1$, rather than only for $y\gg 181$, that in
the region $y\gg 181$ reproduces \eq{phi}, including all the
subleading terms written there, and that in the region $y<10$ where the
numerical integration is possible reproduces very well the 
data.
An unexpected bonus from this analysis will be that we will also find terms
$\sim\exp\{-{\rm const.} y\}$,
that we will relate in sect.~(\ref{renormalons})
to renormalon singularities.

\begin{figure}
\centering
\includegraphics[width=0.7\linewidth,angle=0]{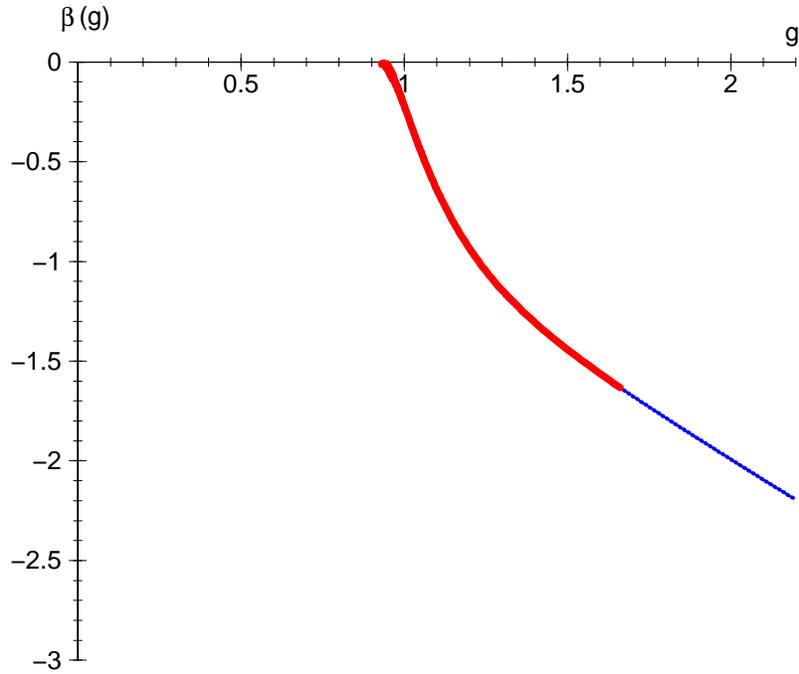}
\caption{The beta function  against $\gym$:
numerical integration  (red solid line), 
 and IR analytic limit (blue, dotted). The UV analytic limit is
$\beta (g)\sim -g^3$, and cannot be distinguished from zero 
on this vertical scale.}
\label{figbeta}
\end{figure}

\begin{figure}
\centering
\includegraphics[width=0.7\linewidth,angle=0]{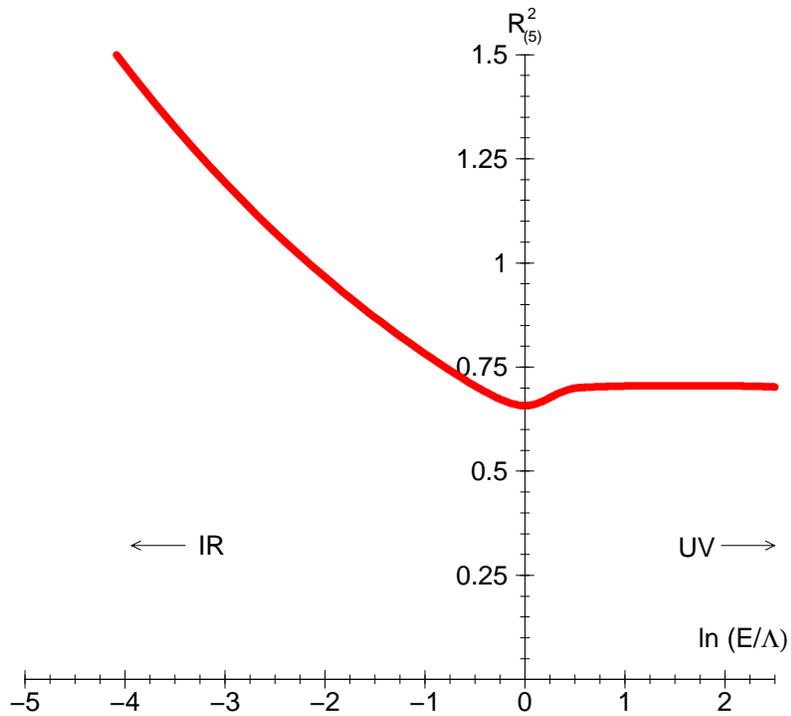}
\caption{$R_{(5)}^2$ against $\ln (E/\Lambda )$. }
\label{figR5}
\end{figure}

The physics of the solution is  illustrated by 
figs.~(\ref{figgym})-(\ref{figR5}).
Fig.~(\ref{figgym}), as already discussed, 
shows $\exp\{ -\phi /2\}$, which is proportional
to the inverse coupling $1/\gym^2$, as a
function of $\ln (E/\Lambda )$. 
This plot can also be used to define the constant
$\Lambda$ which fixes the energy scale. We define it as the value of
$E$ when asymptotic freedom sets in, so by definition in 
fig.~(\ref{figgym}) the change of regime takes place at $\ln
(E/\Lambda ) =0$. On the right we have asymptotic freedom,
since $1/\gym^2\sim\ln (E/\Lambda )$, while on the left side of the plot we
have strong coupling and confinement.

Fig.~(\ref{figbeta}) shows the beta functions $\beta (\gym )$ for our
solution. It has no zero except for the perturbative one
at $\gym =0$, in contrast with the  two-loop result,
which has a second zero at $1/\gym^2 =N/(4\pi)^2$, of course well
outside the limit of validity of the perturbative computation,
since there the 't~Hooft coupling is $N\gym^2=(4\pi )^2\gg 1$. 
The numerical result in the intermediate region
matches well the analytic results at weak and strong 
coupling\footnote{With the vertical scale used in the figure, needed to 
show the matching of the
numerical result with the IR analytic behaviour, the UV behaviour
$\beta (g)\sim -g^3$ cannot be distinguished from the horizontal axis,
but the numerical result indeed matches with $\sim -g^3$ in the UV.
Note also that, to produce this plot, 
we have set to unit the proportionality constant 
in \eq{gym}. Once one has the correct proportionality constant, the
correct figure is obtained with a rescaling of the units of the axes.}.

Another interesting quantity is the  radius of the 5-sphere (in the
Einstein frame, since in this frame the UV metric approaches
$AdS_5\times S^5$) $R_{(5)}$, which is given by \eq{metrica},
\be
R_{(5)}^2=e^{\frac{\xi}{2}-\eta}\, .
\ee
If the adjoint scalar fields become massive, the radius of the
5-sphere  shrinks to zero. We see from fig.~(\ref{figR5}) that 
$R_{(5)}^2$ diverges in the IR limit. At the transition between the IR
and UV region it bounces and finally settles to the constant value
$1/\sqrt{2}$ predicted by \eqs{xi}{eta}. We see that it never
vanishes, and therefore the 6 adjoint scalars remain massless.

\vskip1cm

\vskip1cm

\vskip1cm

\vskip1cm

\vskip1cm

\vskip1cm

\clearpage

\section{Analytic solution}\label{sec4}
In this section we find an analytic expression for the RG flow which
reproduces very well the solution from the IR to the UV regions, 
and even allows to compute exponentially small corrections in the UV
limit. 

The results of the previous section suggest the following 
improved UV ansatz
\bea
e^{-\phi /2}&=& A+by + u(y)\,\label{A} \label{impphi}\\
\xi  (y)&=& -y           +\ln 2 + \zeta (y)\, \\
\eta (y)&=& -\frac{y}{2} +\ln 2 + \omega (y)\, \\
T(y)    &=& -1+t(y)\label{impT}\, .
\eea
where $u, \zeta, \omega$ and $t$ are small, in a sense that will be
clarified below. 
Here $b=2^{-15/2}\simeq 1/181$,  and $A$ is 
a  constant of order one, yet to be fixed ($A>0$
otherwise the solution would not make sense when $by <-A$).
At first, we linearize  the equations of motion
with respect to  $u,\zeta ,\omega$ and $t$ 
(we will discuss later the  non-linear terms), and we get
\bea
\zeta''+\zeta' -2\zeta &\simeq& -\frac{2b}{A+by}
\[ 1+\frac{\zeta}{2}-5\omega-2t-\frac{u}{A+by} \]\, ,\label{linzeta}\\
\omega''+\omega' -2\omega &\simeq& -\frac{2b}{A+by}
\[ 1+\frac{\zeta}{2}-5\omega-2t-\frac{u}{A+by} \]\, ,\label{linomega}\\
t''+t' +2t&\simeq& +\frac{16b}{A+by}
\[ 1+\frac{\zeta}{2}-5\omega-t-\frac{u}{A+by}\]\, ,\label{linT}
\eea
\be\label{linu}
u''+\( 1-\frac{2b}{A+by}\) u'+\frac{b^2}{(A+by)^2}u
\simeq \frac{b^2}{A+by}+b(\frac{\zeta}{2} -5\omega -2t)\, .
\ee
where the prime is $d/dy$. We can now take advantage of the fact that
$b$ is numerically small, and we treat it as a small expansion 
parameter. Instead  $A+by$ in the UV is at least of order one.  Then,
to lowest order in $b$ a particular solution of 
\eqst{linzeta}{linT} is  given simply by 
\be\label{start}
\zeta (y)=\omega (y)=\frac{1}{8}t(y)=
\frac{b}{A+by}\[ 1+O\(\frac{b}{A+by}\)\]\, .
\ee
In fact, in the region where $A+by=O(1)$, $\zeta'$ is smaller by a
factor $b$ compared to $\zeta$ while for $y>>1/b$ we have $1\gg \zeta\sim 1/y
\gg \zeta'$, and similar for $\omega, t$. Substituting \eq{start} into 
\eq{linu} we  obtain the particular solution
\be
u(y) = -\frac{39}{2}b\ln (A+by) \[ 1+O\(\frac{b}{A+by}\)\]\, ,
\ee
To obtain the most general solution, we  first tentatively assume that
the terms $\sim \zeta ,\omega ,t$ and $\sim u/(A+by)$ on the right-hand side of 
\eqst{linzeta}{linu} can be neglected compared to 1. Then the most
general solution is obtained adding to the above particular solution the
solution of the homogeneous equations
\bea
\zeta''+\zeta' -2\zeta &\simeq&0\, ,\hspace{15mm}
\omega''+\omega' -2\omega \simeq0\, ,\nonumber\\
t''+t' +2t&\simeq& 0\, ,\hspace{15mm}\label{hom}
u''+ u'=0\, .
\eea
This would give
\bea
\zeta (y)&\stackrel{?}{=} & \frac{b}{A+by}\[ 1+O\(\frac{b}{A+by}\)\]
+B_{\xi}e^{-2y}  \, ,\label{zeta}\\
\omega (y)&\stackrel{?}{=} &\frac{b}{A+by}\[ 1+O\(\frac{b}{A+by}\)\]
+B_{\eta}e^{-2y} \, ,\label{omega}\\
t (y)&\stackrel{?}{=}& \frac{8b}{A+by}\[ 1+O\(\frac{b}{A+by}\)\]+
B_{T}e^{-y/2} \sin\( \frac{\sqrt{7}}{2}y+\alpha_T \) 
  \, ,\\
u(y)&\stackrel{?}{=}& -\frac{39}{2}b\ln (A+by)\[ 1+O\(\frac{b}{A+by}\)\]
+B_u e^{-y} \, ,
\eea
where $B_{\xi},B_{\eta},B_{T},B_u$ and $\alpha_T$ are arbitrary
coefficients of the homogeneous solutions. Note that the homogeneous
equation for the tachyon has two complex solutions 
$t=\exp\{(-1\pm i\sqrt{7})y/2\}$, 
and then the solution is an oscillating function, as indeed we
expected from fig.~(\ref{figT}).\footnote{We discarded 
exponentially growing solutions of \eq{hom}, as well as the constant
solution of $u''+u'=0$, since the constant part of $u(y)$  has already
been taken care of by the constant $A$, \eq{A}.}

However, if we now insert back $ \zeta ,\omega ,t$ and $u/(A+by)$ 
on the right-hand side of \eqst{linzeta}{linu} to check whether the
approach is self-consistent, we see that the tachyon
gives a contribution 
$\sim bB_Te^{-y/2}\sin ((\sqrt{7}/2)y+\alpha_T)$ 
to the right-hand side of
\eqst{linzeta}{linu}, so that, for instance, 
the term $e^{-2y}$ in the solution for
$\zeta$ dominates over this contribution only if $be^{-y/2}\ll e^{-2y}$
or $y\ll \ln (1/b)$. In the opposite limit
$y\gg \ln (1/b)$ (which with our value of $b$ means $y\gg 3$)
before considering terms $\sim B_{\xi}e^{-2y}$ we must also include the
contribution of the exponentially small term coming from
$t$  to the right-hand side of \eqst{linzeta}{linu}, and we must also
go beyond the linear approximation: e.g., if we want to get correctly all
the terms up to $e^{-2y}$ we must include the contributions from
terms $\sim t^2,t^3, t^4$ and $u^2$. The correct result  then 
turns out to be of the form
\bea
e^{-\phi (y) /2}&=& A+ by-
\frac{39}{2}b\ln (A+by)\[ 1+O\(\frac{b}{A+by}\)\]+ \nonumber\\
&+ &b\, B_Te^{-y/2} \sin\( \frac{\sqrt{7}}{2}y+\alpha_T \) 
\[ 1+O\( b g_1(y)\)\] + \label{solphi}\\
&+&  \[ B_u +bB_T^2g_2(y)\]  e^{-y} +O\( e^{-3y/2}\)
\, ,\nonumber
\eea
with $g_1(y),g_2(y)$ non-singular and oscillating functions of $y$. 
In this work, we are not
particularly interested in their explicit form (for the comparison
with the numerical data these terms play a very marginal role), 
so we  computed explicitly only the
term $\sim e^{-y/2}$, at lowest order in $b$, but it
is important conceptually to understand the structure
of the exponentially small corrections, as we will
see in sect.~(\ref{renormalons}).

A similar structure can be obtained for $\xi ,\eta, T$,
\bea
\label{solxi}
\xi (y)&= & -y +\ln 2+ \frac{b}{A+by}\[
1+O\(\frac{b}{A+by}\)\] \\
& &-\frac{b B_T}{A+by} e^{-y/2} \sin\( \frac{\sqrt{7}}{2}y+\alpha_T \)
\[ 1+O\(\frac{b}{A+by}\)\] + O(e^{-y}) \, ,\nonumber\\
\label{soleta}
\eta (y)&=& -\frac{y}{2}+\ln 2+ \frac{b}{A+by}\[
1+O\(\frac{b}{A+by}\)\]\\
& &-\frac{b B_T}{A+by} e^{-y/2} \sin\( \frac{\sqrt{7}}{2}y+\alpha_T \)
\[ 1+O\(\frac{b}{A+by}\)\] +O(e^{-y})\, ,\nonumber\\
\label{solT}
T (y)&=&-1+ \frac{8b}{A+by}\[ 1+O\(\frac{b}{A+by}\)\]+\\
& & +B_{T}e^{-y/2} \sin\( \frac{\sqrt{7}}{2}y+\alpha_T \)
 \[ 1+O\(\frac{b}{A+by}\)\] +O\( e^{-y}\) \, .\nonumber
\eea
We can now compare with the numerical solution and extract $A,B_T,B_u,\alpha_T$
from a fit. We find $A\simeq 1.1, B_T\simeq 0.8, B_u\simeq -0.6,
\alpha_T\simeq 0.6$.
This analytic solution is shown in 
figs.~(\ref{figT}-\ref{figeta}) (black dashed line)
and we see that it reproduces very well the numerical 
solution already for $y\gsim 1$.

Expanding \eqst{solphi}{solT} in the UV we find that they reproduce
\eqst{phi}{T}, including all subleading terms. The coefficient
$\kappa$ is related to $A$ by 
\be\label{kappa}
\kappa =39\ln b -\frac{2A}{b}= -\frac{585}{2}\ln 2 - 256\sqrt{2}\, A \, ,
\ee
As we noted above, our solution
makes sense only for $A>0$, since otherwise $\ln (A+by)$ becomes
imaginary at intermediate values of $y$. Eq.~(\ref{kappa}) then shows
that there exists a critical value of $\kappa$,
\be
\kappa_c= -\frac{585}{2}\ln 2
\ee
and that our solution (\ref{solphi})-(\ref{solT}) can produce, in the
deep UV region $y\gg 1/b$, only solutions of the form 
(\ref{phi})-(\ref{T}) with $\kappa <\kappa_c$. 
The solution considered by Klebanov and Tseytlin~\cite{KT2} has instead
$\kappa =0$ and it does not belong to this class. Indeed, ref.~\cite{KT2}
finds that in the IR it connects to another conformal fixed point at
infinite coupling, rather than to a confining solution.

We see therefore that $\kappa$ plays a crucial role in determining
the IR behaviour of the RG flow. In the language of Wilson's
renormalization group, this means that $\kappa$ is a parameter that
determines to which universality class  the UV theory belongs, and we
have found that there are (at least) two universality classes:
for $\kappa <\kappa_c$ we are in the domain of attraction of a
confining IR fixed point, while for $\kappa >\kappa_c$ we are in the
domain of attraction of a different fixed point.

It is also important to understand how the result changes if we change the
form of the function $f(T)$. We find that, as long as $f(T)$ has a
minimum at a finite value $T_0$, the qualitative behaviour of the
solution does not change. However, it is especially
important to known whether
the exponential terms $\sim e^{-y/2}$ are affected by the form of
$f(T)$. Linearizing \eq{d} for $f(T)$ generic, and writing
$T=T_0+t$,  the left-hand side of \eq{linT} becomes
$t''+t'+c t$, with $c=f''(T_0)/(2f^2(T_0))$; for $f(T)=1+T+T^2/2$,
$c=2$. The homogeneous equation now has the solutions $t=e^{\gamma y}$
with
\be
\gamma =-\frac{1}{2} (1\pm i\sqrt{4c-1})\, .
\ee
We see that, as long as $c\geq 1/4$, Re~$\gamma$ is unchanged  and we
still have solutions of the form $e^{-y/2}\sin (\nu y)$, but the
frequency $\nu$  depends on the explicit form of $f(T)$. Therefore the value
that we found, $\nu =\sqrt{7}/2$, does not have a special
significance, being related to the unknown form of $f(T)$, while the factor
$e^{-y/2}$ is universal as long as $f(T)$ obeys
\be\label{f''}
2f''(T_0)\geq f^2(T_0)\, .
\ee
The function $g(T)$ instead does not enter the homogeneous equation,
so from this point of view its form is irrelevant.

Finally, it is straightforward to compute analytically 
the corrections to the
solution in the IR regime. The corrections are exponentially small in
$\rho$,
\bea
\phi  = &\phi_1\rho+\phi_0& - \sum_{i=1}C_i^{(\phi )} 
\exp\{-\gamma_i^{(\phi )}\rho \}\, ,\nonumber\\
\xi  = &\xi_1\rho+\xi_0 &- \sum_{i=1}C_i^{(\xi )} 
\exp\{-\gamma_i^{(\xi )}\rho \}\, ,\label{IRsub}\\
\eta  = &\eta_1\rho+\eta_0& - \sum_{i=1}C_i^{(\eta )} 
\exp\{-\gamma_i^{(\eta )}\rho \}\, ,\nonumber\\
  &T=  & -\sum_{i=1}C_i^{(T )}  \exp\{ -\gamma_i^{(T )} \rho\}\nonumber\, .
\eea
We have computed the coefficients $C_i,\gamma_i$ for $i=1,2,3$. We get,
for $\phi_1=\xi_1=\sqrt{5}\eta_1$,
\begin{eqnarray}
\gamma_1^{(\phi)} = (\sqrt{5} + 3) \phi_1 \qquad  \gamma_2^{(\phi)} 
& = & 2 (\sqrt{5} + 1) \phi_1  \qquad \gamma_3^{(\phi)} = 
\frac{9 + 3 \sqrt{5}}{\sqrt{5}} \phi_1 \nonumber\\
\gamma_1^{(\xi)} = 2 \phi_1 \qquad \gamma_2^{(\xi)} & = & 4 \phi_1 \qquad 
\gamma_3^{(\xi)} = (\sqrt{5} + 3) \phi_1 \nonumber\\
\gamma_1^{(\eta)} = \frac{4}{\sqrt{5}} \phi_1 \qquad \gamma_2^{(\eta)} & = & 
  \frac{8}{\sqrt{5}} \phi_1 \qquad  \gamma_3^{(\eta)} = (\sqrt{5} + 3) \phi_1 \nonumber\\
\gamma_1^{(T)} = 2\phi_1 \qquad \gamma_2^{(T)} & = & (\sqrt{5} +1) \phi_1 \qquad \gamma_3^{(T)} = 4\phi_1\nonumber
\end{eqnarray}
\begin{eqnarray}
C_1^{(\phi)} & = & \frac{1}{16(\sqrt{5} +3)^2 \phi_1^6} e^{\frac{1}{2}\phi_0 -\frac{7}{2}\xi_0  -5\eta_0} \qquad 
C_2^{(\phi)}  = - \frac{1}{16(\sqrt{5} + 1)^4 \phi_1^8} e^{\phi_0 - 3\xi_0 -10\eta_0} \nonumber\\
 C_3^{(\phi)} & = & \frac{125}{32 (9 + 3\sqrt{5})^2 \phi_1^8} e^{\frac{1}{2}\phi_0 - \frac{7}{2}\xi_0 -9\eta_0} \qquad
C_1^{(\xi)} = \frac{1}{2 \phi_1^2} e^{-2\xi_0} \nonumber\\
C_2^{(\xi)} & = & \frac{3}{16 \phi_1^4} e^{-4\xi_0} 
\qquad C_3^{(\xi)} =- \frac{1}{(\sqrt{5} + 3)^2 \phi_1^6} 
\left(\frac{2}{(\sqrt{5} + 1)^2} - \frac{1}{16}\right)
 e^{\frac{1}{2}\phi_0 -\frac{7}{2}\xi_0 -5\eta_0 }\nonumber\\
C_1^{(\eta)} & = & \frac{5}{2 \phi_1^2} e^{-4\eta_0} \qquad C_2^{(\eta)} =  \frac{25}{4\phi_1^4} e^{-8\eta_0}  \nonumber\\
C_3^{(\eta)} & = & 
\frac{1}{16(\sqrt{5} +3)^2 \phi_1^6} e^{\frac{1}{2}\phi_0 -\frac{7}{2}\xi_0  -5\eta_0 }\qquad
C_1^{(T)} = \frac{1}{2 \phi_1^2} e^{-2\xi_0} \nonumber\\ 
C_2^{(T)} & = &- \frac{1}{(\sqrt{5} + 1)^2 \phi_1^4} e^{\frac{1}{2}\phi_0 -\frac{3}{2}\xi_0 -5\eta_0 } \qquad C_3^{(T)} = \frac{3}{16 \phi_1^4} e^{-4\xi_0}\nonumber
\end{eqnarray}
This analytic solution is shown  in figs.~(\ref{figT}-\ref{figeta})
as a blue dotted line, and
we see that it reproduces very well the numerical solution
in the IR region. Having included the  the first three subleading
exponential, we see that the IR
analytic solution has an overlap with the analytic UV
solution, so that the two expansions together describe analytically
the solution in the whole region.

\section{The renormalon singularities}\label{renormalons}

A striking feature of the UV solution  is the appearance of
the exponentially small terms $\sim e^{-y/2}, e^{-y}$, etc.; in the
UV, at leading order, $y\simeq -\xi =4\ln E$, so (extracting an
overall proportionality factor $4b$)  \eq{solphi} gives
\be\label{final}
\frac{1}{\gym^2(E)}\sim \ln \frac{E}{\Lambda}-\frac{39}{8}\ln\ln \frac{E}{\Lambda}
+\frac{B_T}{4}F_1(E) \frac{\Lambda^2}{E^2} +B_T^2F_2(E) \frac{\Lambda^4}{E^4} +
C \frac{\Lambda^4}{E^4}+
O(\frac{1}{E^6})
\ee
with 
\be\label{F1}
F_1(E)= \sin\( \beta_T\ln \frac{E}{\Lambda}+\alpha_T \) 
\, \[ 1+ O(b) \] \, ,
\ee
and $C=B_u/(4b)$ (however, we expect that the numerical values of these
coefficients will be affected by our uncertainties, as it happens to
the factor $39/8$).
$F_2(E)$ can be determined adding the terms $\sim t^2$ to \eqs{linzeta}{linu}.
The positive constant $\beta_T$ is given by
\be
\beta_T=2\[ \frac{2f''(T_0)}{f^2(T_0)} -1\]^{1/2}\, ,
\ee
and depends on the explicit form of $f(T)$, so  its numerical value is not
under control. The exponential correction to the relation $y\simeq -\xi$
gives an $O(b)$ contribution to (\ref{F1}).

First of all, we read from \eq{final} that there are
power corrections, and they are  exactly of the form that
is expected from the effect of renormalons. 

Let us recall that, given a divergent series
\be
G(\gym^2)=\sum_{n=1}^{\infty}G_n\gym^{2n}\, ,
\ee
with $|G_n|$ bounded by $r^n n!$ at large $n$,
one defines its Borel transform $B(z)$ as
\be
B(z)=\sum_{n=0}^{\infty}\frac{G_{n+1}}{n!} z^n\, ,
\ee
inside its convergence radius, $|zr|<1$, and by analytic continuation elsewhere.
Then
\be\label{Borel}
G(\gym^2)=\int_0^{\infty}dz\, e^{-z/\gym^2}B(z)\, 
\ee
(where the integration contour is on the positive real axis)
can provide a resummation of the perturbative series, depending on the
structure of the singularities of $B(z)$ in the complex  plane
(Borel plane), and on the convergence properties of the integral at infinity.
For asymptotically free non abelian gauge theories
a number of singularities in the Borel plane
have been identified as follows~\cite{Par,tH1,Alt,Ben}.
IR renormalons give singularities on the positive semiaxis, at
\be
\bar{z}=\frac{2}{\beta_0}k\, ,\hspace{15mm} k=2,3,\ldots
\ee
($\beta_0>0$ for asymptotic freedom; in our case
$\beta_0=\frac{8}{3}N/(8\pi^2)$, see \eq{beta0}).
Since these singularities are on the positive semiaxis, we 
must supplement \eq{Borel} with a prescription for
dealing with them, and this produces 
non-perturbative contributions of order
\be\label{power}
\exp\{ -\frac{2k}{\beta_0\gym^2} \}\sim \frac{1}{E^{2k}}\, ,
\ee
times regular functions of the energy.
Since the first IR renormalon is at $k=2$, it gives a contribution
$\sim 1/E^4$. UV renormalons are instead on the negative semiaxis, at
\be
\bar{z}=-\frac{2}{\beta_0}k\, ,\hspace{15mm} k=1,2,3,\ldots
\ee
Even if these singularities are not on the integration contour, they
limit the convergence radius of the expansion for $B(z)$ and 
are responsible for effects  of the form 
\be
\exp \{ -\frac{|\bar{z}|}{\gym^2} \}
\sim \frac{1}{E^{2k}}
\ee
However, now
$k$ starts from 1, so that the leading effect is $1/E^2$.
There are further  singularities on the positive
semiaxis due to instantons, but their effect is of order
$\exp (-8\pi^2/\gym^2)$, which for $SU(N)$ with 6 adjoint scalars is
$\sim (1/E)^{8N/3}$ (or $\sim (1/E)^{16N/3}$ if the first
non-vanishing contribution comes from an instanton-antiinstanton pair)
 and therefore is quite suppressed for large $N$.

Comparing our result (\ref{final}) with the above situation, we see that the
$1/E^2$ term that we have found matches with the contribution of the
first UV renormalon. The fact that we get the correct location of the
renormalon singularity in the Borel plane is  non-trivial, and
can be traced back to the fact that the linearization of \eq{d} gave
\eq{linT}, which has the homogeneous solution $t(y)=e^{\gamma y}$ 
with Re~$\gamma =-1/2$, as long as \eq{f''} holds.
From the point of view of the gravity
computation, we could have obtained $1/E^{\alpha}$ with
$\alpha$ {\em a priori} any real number, not 
necessarily integer nor rational.
We therefore regard this agreement as a successful and non
trivial test of the conjectured duality between type 0B theory and non
supersymmetric YM theory.

The term $B_T^2F_2(E)/E^4$ comes from the iteration of the term
$B_TF_1(E)/E^2$, as we see from the coefficient  $B_T^2$,  so it clearly
corresponds to the UV renormalon with $k=2$.  Iterating further the
contribution of the tachyon  in the solution of the equations of
motion, we find all the UV renormalons at their correct locations in
the Borel plane.

The term $C/E^4$ instead has a prefactor which is just equal to the
constant $C$ and is not related to $F_1(E)$ or $B_T$,  so 
it has a different physical origin. 
Thus, we identify it with the $k=2$ IR renormalon.
Again, it is very remarkable that its position in the Borel
plane  matches exactly the position of
the first ($k=2$) IR renormalon; furthermore, we correctly find no
term corresponding to an IR renormalon with $k=1$, which corresponds
to the absence of gauge-invariant operators of dimension 2 in the OPE
of current-current correlation functions.

Finally, even the energy dependence of the prefactors is consistent
with what is known about renormalons from the analysis of Feynman
graphs. In fact, 
in the approximation in which only a single chain of bubbles is
considered, it is known that all UV renormalons are double poles,
while the $k=2$ IR renormalon is a single pole, and all others IR
renormalons are again double poles~\cite{Ben}. For a single pole 
\be
B(z)\sim \frac{1}{z-\frac{4}{\beta_0}},
\ee
and then \eq{Borel} gives a contribution $\sim \exp\{- 4/(\beta_0\gym^2)\}\sim
1/E^4$ without any further energy dependence, in full agreement with
our interpretation of $C/E^4$ as the contribution of the $k=2$ IR
renormalon.
For a double pole on the positive semiaxis, 
instead, the non-perturbative contribution is of order
\bea
G(\gym^2)&\sim &\int_0^{\infty}dz\, e^{-z/\gym^2}\frac{1}{(z-\bar{z})^2}
\ra \nonumber\\
&\ra& \frac{1}{\gym^2}\int_0^{\infty}dz\,
e^{-z/\gym^2}\frac{1}{(z-\bar{z})}\sim \frac{1}{\gym^2}e^{-\bar{z}/\gym^2}\, ,
\eea
where in the second line we have integrated by parts and kept only the
non-analytic term. This provides a $\ln E$ term in the prefactor.
Similarly, there are logarithmic terms from the
double poles on the negative semiaxis~\cite{Ben,BZ}. Thus
in the single chain approximation 
the first UV renormalon is  of order
\be
\sim \frac{\beta_0\ln E}{E^2}\, ,
\ee
and in our approach is reproduced
by the first term of the expansion of \eq{F1} in
powers of $\beta_T$ (which suggests that $\beta_T$ is proportional to
$\beta_0$). Higher order terms are expected to come from
graphs with multiple chains insertions, and are not well understood in
QCD~\cite{Ben}. In this sense, \eq{F1} is also a remarkable prediction
of the dual string theory, that would be very interesting to check
from the YM side.

\begin{figure}
\centering
\includegraphics[width=0.7\linewidth,angle=0]{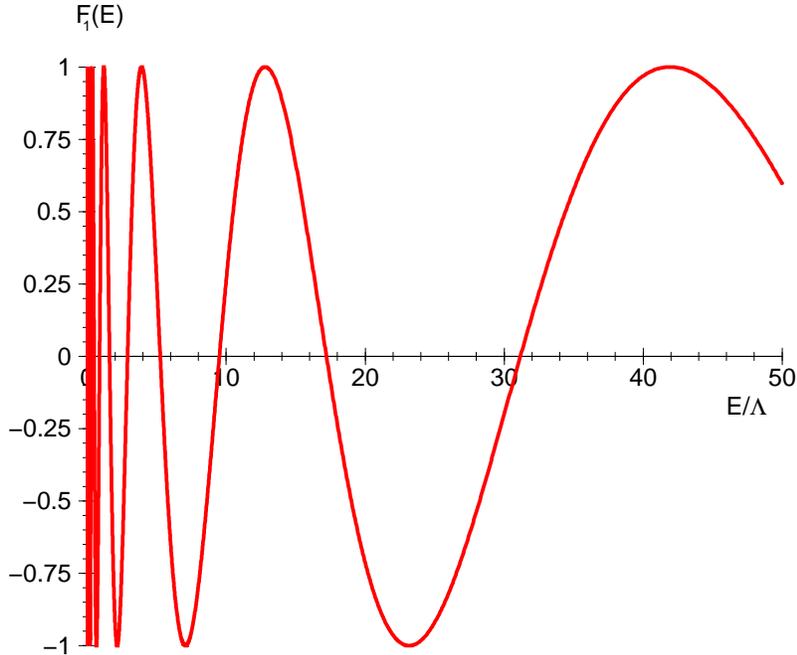}
\caption{The function $ F_1(E)$. }
\label{figF1}
\end{figure}

At low energies $F_1(E)$  oscillates faster and faster, 
see (fig.~\ref{figF1}). This is physically very reasonable, since
these oscillations signal the transition from asymptotic
freedom to  the confining regime. Thus, the form of $F_1(E)$ seems to
capture both the correct single chain result and a physically correct
behaviour when the  IR is approached.

We conclude this section with a general comment.
The  rule concerning the position of renormalons in field
theory is that  we find on the positive semiaxis the renormalons
corresponding to strongly coupled physics. For
asymptotically free gauge theories
 IR renormalons are at $z>0$, while UV renormalons are at $z<0$. 
In QED or $\lambda \phi^4$
the coupling becomes strong in the UV and correspondingly the UV
renormalons are at $z>0$ and IR renormalons are at $z<0$.
When a singularity is at $z>0$ we must give a prescription for going
around it; if we understand the physics associated to this
singularity,  we can give a physically motivated prescription; then we
are left with a non-perturbative contribution, but no ambiguity on the
resummation of the perturbative series. This is the case of
instantons~\cite{tH1}. If however we do not understand the physics
behind the singularity, we do not know what prescription to use.
Different prescriptions (e.g. going around a pole from above or from
below) give answers that differ by terms of order $\exp
(-2k/\beta_0\gym^2)$, so renormalons on the positive
semiaxis are usually taken as a source of
an uncertainty of this order in the 
resummation of the perturbative 
series\footnote{Actually, in QCD even UV renormalons  can give rise to 
uncertainties on the perturbative series which from the Feynman graphs
point of view are  uncalculable, despite the fact that the UV physics is
well understood~\cite{VZ}.}.

In our case, however, the coefficients $B_T, C$ 
in \eq{final} are fixed  from the comparison
with the numerical solution. Once we specify the
IR solution, the values of
$B_T,C$, etc. follow, and the renormalon contribution, rather than an
uncertainty on the perturbative series, becomes a well defined
non-perturbative contribution. 
The fact that IR physics fixes the renormalon contributions
confirms the old conjecture of
't~Hooft~\cite{tH1} that renormalon singularities are related to the
quark confinement mechanism. It is  interesting to note that instead
$B_T,C$ cannot be fixed, even in principle, from the UV side, 
since there they just appear
as integration constants of the homogeneous equations~(\ref{hom}).

\section{$\alpha'$ and string loop corrections}\label{alpha'}

We next consider the effect of $\alpha'$ corrections on the
solutions. For the type~0 theory, these have been discussed
in ref.~\cite{KT1}, where it is found that they have the same structure as
in type~II theories; in particular, world-sheet supersymmetry implies
the vanishing of $\sim\alpha' ({\rm Riemann })^2$ corrections, and the first
non-vanishing correction is $\sim\alpha'^3 ({\rm Riemann })^4$. It is
well known that $\alpha'$ corrections suffer from a certain degree of
ambiguity, because not all the coefficients of the possible operators
are fixed by the comparison with the string amplitudes; a related
ambiguity is the fact that one can perform field redefinitions that
mix different orders in $\alpha '$, e.g. $g_{\mu\nu}\ra g_{\mu\nu}+
\alpha 'R_{\mu\nu}$. If one has an exactly conformal background these
field redefinitions do not matter, but of course if one works at a
finite order in the $\alpha '$ expansion they make a difference.

An important observation~\cite{KT2} is that we can choose the
$\alpha'$ correction such that they depend only on the Weyl tensor,
rather than  on the Riemann tensor, and then they have a much milder
behaviour when we approach $AdS_5$. To be quantitative, consider
\eq{a}, which, including the first non-vanishing $\alpha '$
correction, has the general
form~\cite{KT2} 
\be
\ddot\phi+\frac{1}{2} g(T) e^{\frac{\phi+\xi}{2}-5\eta}=
c\,  e^{\frac{-3\phi+\xi}{2}-5\eta}\, ({\rm Weyl })^4\, ,
\ee
where $c$ is a constant, (Weyl)$^4$ denotes the appropriate contractions
of four Weyl tensors $C_{MNRS}$, and again we have set $\alpha '=1$.
Writing $\ddot\phi =e^{2y}(\phi''+\phi')$, where as before 
the dot is $d/d\rho$ and the prime
is $d/dy$, with $\rho =e^{-y}$, we have
\be\label{phiW}
\phi''+\phi'+\frac{1}{2} g(T) e^{\frac{\phi+\xi}{2}-5\eta-2y}=
c\,  e^{\frac{-3\phi+\xi}{2}-5\eta-2y}\, ({\rm Weyl })^4\, .
\ee
We see  that an ansatz of the form (\ref{phi})-(\ref{T}), or its
improved form  (\ref{impphi})-(\ref{impT}), is still a solution, but
the numerical coefficients in front of the factors $\ln 2$ change. In
fact, consider the ansatz $\phi =-2\ln y
+a_1\ln 2 +O(\ln y/y), \xi =-y+a_2\ln 2+O(1/y),
\eta =-y/2+a_3\ln 2 +O(1/y), T=-1+O(1/y)$. 
We have checked  that on this ansatz
all components of the Weyl tensor $C_{MNRS}$ are $O(1/y)$,
so that~\cite{KT2}
\be
({\rm Weyl})^4\sim\frac{1}{y^4}\, .
\ee
Substituting into \eq{phiW} we get (setting again $g(T)=T^2/2$)
\be
-\frac{2}{y}+\frac{1}{4y}\, e^{(\frac{a_1+a_2}{2}-5a_3)\ln 2}
+O(\frac{1}{y^2})=
\frac{\rm const.}{y}\, e^{(\frac{-3a_1+a_2}{2}-5a_3)\ln 2}
+O(\frac{1}{y^2})\, ,
\ee
and similarly for the equations for $\xi ,\eta ,T$. Thus, the leading
$y$ dependence cancels, and we can adjust the coefficients
$a_1,a_2,a_3$ so that the equations are satisfied. We see that the
numerical values of the coefficients $a_1,a_2,a_3$ are affected by the
inclusion of the (Weyl)$^4$ term. The logarithmic behaviour in the UV is
not spoiled, but the numerical values of the beta function coefficients
change~\cite{KT2}. 

We now want to see whether the terms related to renormalons
are affected. If we use again the improved ansatz
(\ref{impphi})-(\ref{impT}), where now $b$ and the coefficients of 
$\ln 2$ in $\xi ,\eta$ are modified according to the discussion above, we
see that, in the limit $by\gg 1$, where $b/(A+by)\simeq 1/y$, the net
effect of the addition of the (Weyl)$^4$ term is to change  the
numerical factors
inside the square brackets on the right-hand side of 
\eqst{linzeta}{linu}. However, the associated homogeneous equations are
still given by \eq{hom}, and are unaffected by the inclusion of 
the (Weyl)$^4$ term. Therefore, the position of the renormalon
singularities in the Borel plane is unaffected by these $\alpha'^3$
corrections.
It is clear that the argument goes through at all orders in $\alpha'$
as long as the corrections can be written in terms of the Weyl
tensor. As for the string loop corrections, in the UV they
are small by definition, since $e^{\phi}\ll 1$. Thus, it appears that
the result on the position of renormalon singularities is quite robust.

In the IR the situation is more complicated, first of all because
unavoidably when $\gym$ becomes strong also the string loop corrections
become important. This is different from the standard AdS/CFT
correspondence, where the dilaton is constant and $g_s=e^{\phi}$ 
can be taken to be small everywhere.

\begin{figure}
\centering
\resizebox{\textwidth}{6cm}{{\includegraphics{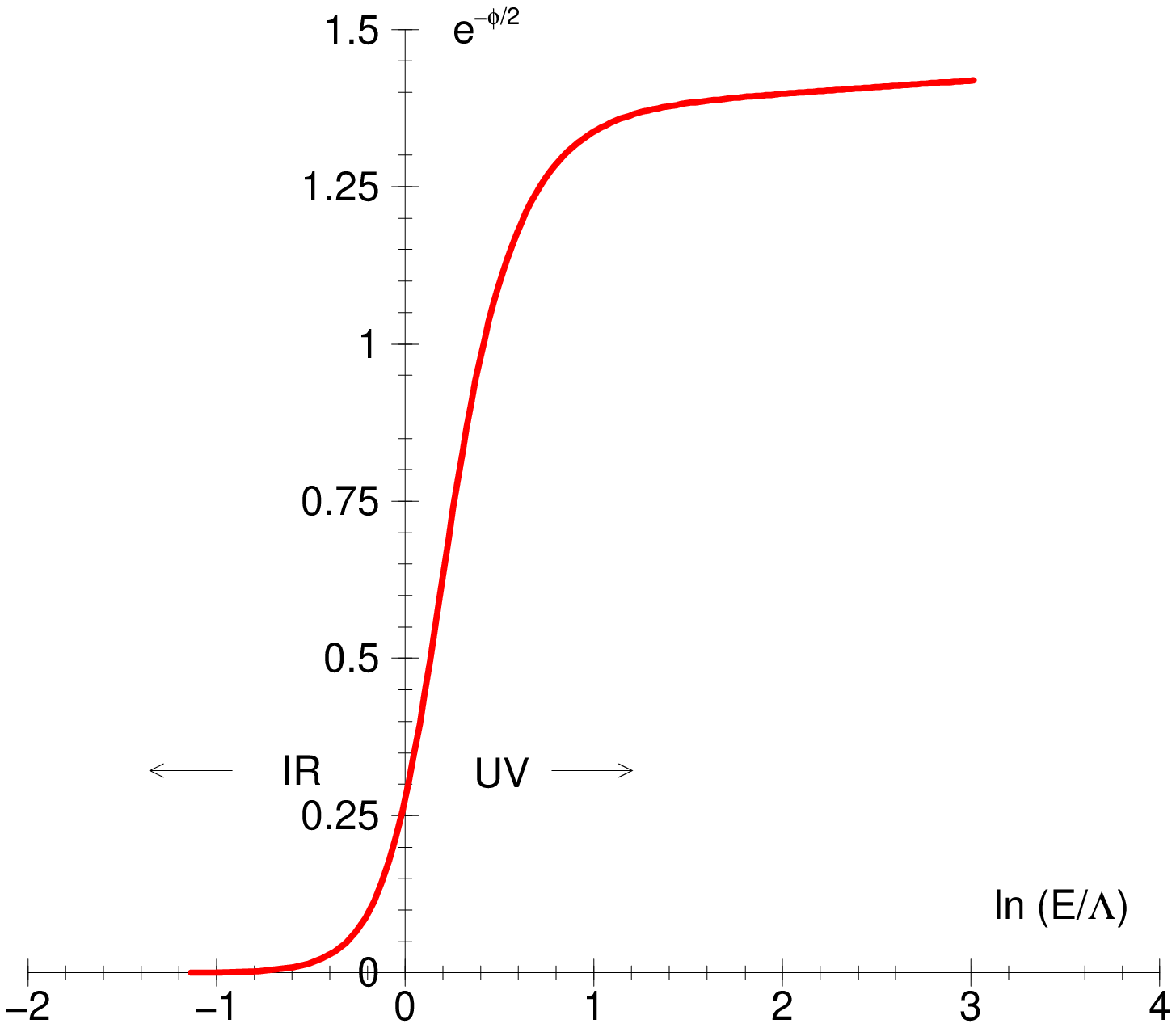}}
{\includegraphics{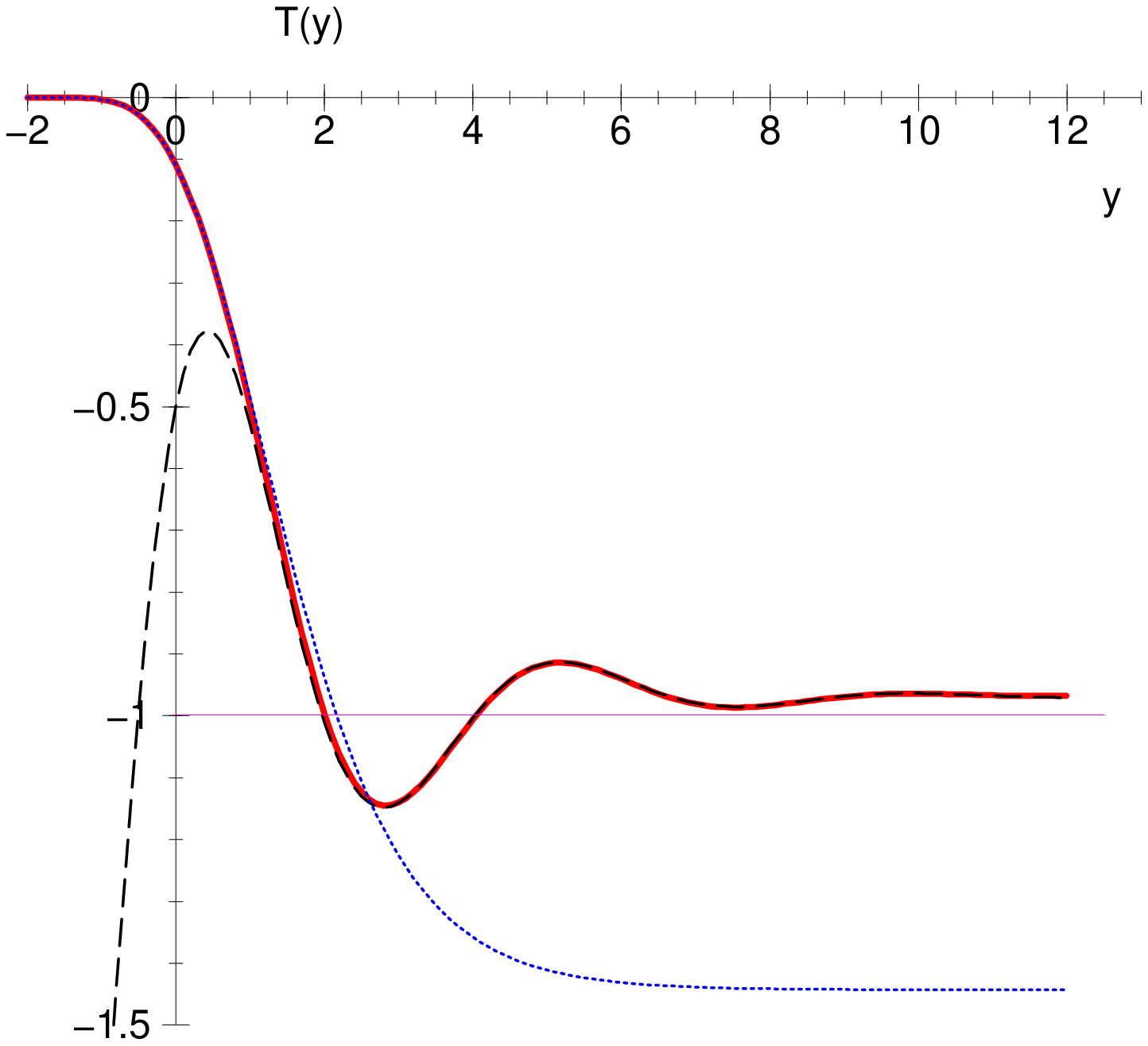}}} 
\caption{$e^{-\phi /2}$ and the tachyon in the case $\phi_1 =3,\xi_1=1$.}
\label{phi3}
\end{figure}

The role of $\alpha '$ corrections in the IR has been discussed in 
ref.~\cite{Min2}, and it has been  found that they depend crucially on
the numerical relations between the coefficients $\xi_1,\phi_1,\eta_1$ 
that parametrize the IR solution (\ref{conf}). With the choice $\xi_1
=\phi_1$ that we used in sect.~\ref{nume}, 
the corrections are indeed large in the
IR, while, for instance, for $\phi_1=3\xi_1, t_1=0$ we still have a confining
solution but now the $\alpha '$ corrections in the IR are
exponentially small in $\rho$. Therefore we have repeated the
numerical analysis discussed in sect.~\ref{nume}, 
for the case $\phi_1=3\xi_1,
\xi_1=1, t_1=0$
and, from \eq{isu}, $\eta_1=1$.
We have found that the lowest order solution for the dilaton and the
tachyon shows no qualitative
difference, see fig.~(\ref{phi3}).

\begin{figure}
\centering
\includegraphics[width=0.6\linewidth,angle=0]{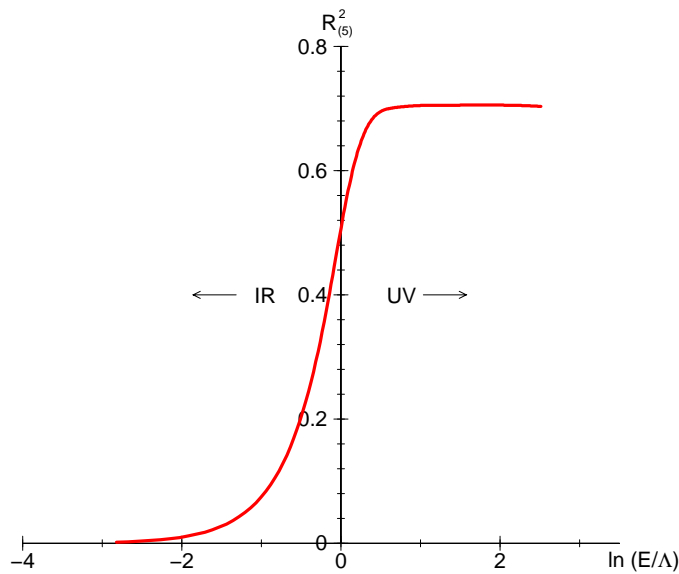}
\caption{$R_{(5)}^2$ in the case   $\phi_1 =3,\xi_1=1$.}
\label{figR5-3}
\end{figure}

The functions $\eta ,\xi$ are also qualitatively very similar
to the previous case;
the main difference is that, since now $\xi_1/2-\eta_1 <0$,
the radius of the 5-sphere
$R_{(5)}^2=e^{\frac{\xi}{2}-\eta}$ goes to zero in the IR, see
fig.~(\ref{figR5-3})
(using \eq{isu}, we see that this happens in general if
$\phi_1^2>(3/2)\xi_1^2$). This  suggests that the six adjoint
scalars become massive, but to clarify this point we need a
better understanding of string loop corrections in the IR limit.

\clearpage

\bigskip


\end{document}